# The Understanding of Intertwined Physics: Discovering Capillary Pressure and Permeability Co-Determination

Omar Alfarisi, Djamel Ouzzane, Mohamed Sassi, and TieJun Zhang

**Abstract**

Although capillary and permeability are the two most important physical properties controlling fluid distribution and flow in nature, the interconnectivity function between them was a pressing challenge. Because knowing permeability leads to determining capillary pressure. Geodynamics (e.g., subsurface water, CO2 sequestration) and organs (e.g., plants, blood vessels) depend on capillary pressure and permeability. The first determines how far the fluid can reach, while the second determines how fast the fluid can flow in porous media. They are also vital to designing synthetic materials and micro-objects like membranes and micro-robotics. Here, we reveal the capillary and permeability intertwined behavior function. And demonstrate the unique physical connectors: pore throat size and network, linking capillary pressure and permeability. Our discovery quantifies the inverse relationship between capillary pressure and permeability for the first time, which we analytically derived and experimentally proved.

## Introduction

Capillary pressure ($P_c$ ) controls the distribution of fluids in natural and synthetic homogeneous and heterogeneous porous media (*1-3*), while permeability controls the fluid flow velocity (*4-6*). Homogeneous structure Permeability represents the flow of a single geometry (*7-9*) and heterogeneous media permeability aggregates the flow of multiple geometries (*10*). Compared to capillary pressure, measuring permeability is faster, easier, and non-destructive for many sample sizes. Therefore, quantifying capillary pressure from permeability is an attractive approach for scientists and engineers. Several methods have been developed in the past to determine the $P_c$. One of the early ones is the J-function (*11*). This function continued to be the primary function for $P_c$ curve determination, although it was for Clastic rather than Carbonate. The limitation of J-function comes from the assumption of having a bundle of capillary tubes (*12-14*). Later an introduction to a link between permeability and $P_c$ emerged with the C-Function.

Two Pc curve points established the connection with permeability. These two points are Transition Zone Top (TZT) and Transition Zone Plateau (TZP), derived from a Cretaceous carbonate formation and validated in another Cretaceous carbonate formation (*13*). However, the C-function is uneasy about the fluid saturation related to these two points. The $P_c$ curves are either for single, dual, or triple Pore Throat Network (PorThN). With C-Function, geoscientists created $P_c$ curves for each permeability value of the single PorThN and generated correlation curves. But these curves are not readily available. Alternatively, the $P_c$ curves held a 45-degree line on the $P_c$ and $S_w$ cross-plot (*13*) and then used a power function for determining the $P_c$ curve. We propose direct quantification of $P_c$ from permeability, an approach we call the Capillary Pressure Permeability Function (P-Function). We analytically derive and experimentally prove the intertwined behavior of these two physical properties.

**Novel $P_c$ Determination; the P-Function**

In this research, we modified TZT and TZP of the C-Function (*13*) to replace it with the P-Function. In addition, the P-Function introduces a direct inverse relationship between $P_c$ and permeability.

We started by the $P_{c.}$ "developed by Plateau and applied to porous media by Leverett" (*15*) as shown in Eq.1 below:

$$P_c = \frac{2\,\sigma\, cos\theta}{r} \qquad (1)$$

Where; $r$ is the pore throat radius, which we also donate $r_{PorTh}$,

$\sigma$ is Interfacial Tension,

$\theta$ is the Contact Angle.

We consider the permeability $k_{3D_{rhombohedral}}$ (*16*) which is the permeability of nature's preferable grain configuration, the Rhombohedral structure. The grain size controls the permeability, and for the existence of more than a grain size in the rock fabric, the parallel and serial aggregation of permeability for each grain size using Morphology Decoder (*16*) determines the heterogeneous permeability:

$$k_{3D_{rhombohedral}} = PorTS_{rhombohedral_{3D_{Effective}}} = 0.0858 r_g^2 \qquad (2)$$

Where; $PorTS_{rhombohedral_{3D_{Effective}}}$ is the area of the pore throat for grain size,

$r_g$ is the grain size.

We can notice that $r_{PorTh}$ defines the $P_c$ in Eq. 1, while $r_g$ defines $k_{3D_{rhombohedral}}$ in Eq. 2. To establish the direct link between $P_c$ and $k_{3D_{rhombohedral}}$ we convert the $r_g$ to $r_{PorTh}$. We replaced 0.0858 with an equivalence value of $0.02731\pi$ for the permeability of Eq. 2 as per Eq. 3 below:

$$k_{3D_{rhombohedral}} = 0.02731\pi \, r_g^2 \tag{3}$$

Now we can equate the right side of Eq. 3 with the equivalent pore throat area of Eq. 2., leading to the Eq. 4 below:

$$\pi \, r_{PorTh}^2 = 0.027311\pi \, r_g^2 \tag{4}$$

Now we simplify Eq. 4 to solve for $r_{PorTh}$ to get the following:

$$r_{PorTh} = 0.027311 \, r_g \tag{5}$$

*Also,*

$$r_{PorTh}^2 = 0.027311 \, r_g^2 = \left(\frac{1}{36.615283}\right) r_g^2 \tag{6}$$

Then by substituting Eq. 6 in Eq. 2, we get Eq. 7 below:

$$k_{3D_{rhombohedral}} = (0.0858).(36.615283) \, r_{PorTh}^2 = 3.14159 \, r_{PorTh}^2 \tag{7}$$

We can now substitute the value 3.14159 in Eq. 7 with ~$\pi$ to have the following equation:

$$k_{3D_{rhombohedral}} = \pi\, r_{PorTh}^2 \qquad (8)$$

Also, we write Eq. 8 for $r_{PorTh}$ to be as shown in Eq. 9:

$$r_{PorTh} = \sqrt{\frac{k_{3D_{rhombohedral}}}{\pi}} \qquad (9)$$

Then to produce the 1st point in the $P_c$ curve, the Displacement Capillary Pressure ($P_{cd}$) point. We substitute Eq. 9 in Eq. 1 to get eq. 10, which we call the P-Function or (the $P_c$ Permeability Function):

$$P_{cd} = 2\frac{\sigma \cos(\theta)}{\sqrt{\frac{k_{3D_{rhombohedral}}}{\pi}}} \qquad \textit{(P-Function or 10)}$$

We used Eq. 8-10 to produce Table-1 to validate these equations with reference experiments data for permeability and grain size (*17*), using parameters ($\sigma$ , $\theta$) for Mercury (Hg), Air, Oil, and Water (*18*). We can see that the $P_c$ reference experiment data matches the calculated $P_c$ using the P-Function for a Cretaceous carbonate rock.

Table-1. Validation of Eq. 8-10 with reference experiments.

| Reference r_Grain (um) Experiment | Derived r_PorTh (um) using r_PorThN-MorphologyDecoder | Reference Experimental Permeability (mD) | K_3DRhombohedral (mD) using r_Grain | K_3DRhombohedral (mD)_using r_PorTh | pc_P-Function (psi) for Hg/air | Reference Pc (psi) for Hg/air |
|---|---|---|---|---|---|---|
| 31 | 5.77 | 100.00 | 103.79 | 104.75 | 9.69 | |
| 41.5 | 7.73 | 210.00 | 186.00 | 187.73 | 7.24 | |
| 62.5 | 11.64 | 420.00 | 421.88 | 425.78 | 4.80 | |
| 88.5 | 16.48 | 830.00 | 845.88 | 853.72 | 3.39 | |
| 125 | 23.28 | 1700.00 | 1687.50 | 1703.13 | 2.40 | |
| 175 | 32.60 | 3300.00 | 3307.50 | 3338.13 | 1.72 | |
| 250 | 46.57 | 6600.00 | 6750.00 | 6812.51 | 1.20 | |
| | | | | 1000 | 3.14 | |
| | | | | 500 | 4.43 | |
| | | | | 400 | 4.96 | |
| | | | | 396 | 4.98 | 5.01 |
| | | | | 250 | 6.27 | |
| | | | | 200 | 7.01 | |
| | | | | 150 | 8.10 | |
| | | | | 100 | 9.91 | |
| | | | | 80 | 11.08 | |
| | | | | 60 | 12.80 | |
| | | | | 55 | 13.37 | 12.52 |
| | | | | 40 | 15.68 | |
| | | | | 30 | 18.10 | |
| | | | | 20 | 22.17 | |
| | | | | 11.6 | 29.11 | 30.85 |
| | | | | 10 | 31.35 | |
| | | | | 8 | 35.05 | |
| | | | | 5 | 44.34 | |
| | | | | 3 | 57.24 | |
| | | | | 1.4 | 83.79 | 87.04 |
| | | | | 1 | 99.15 | |
| | | | | 0.5 | 140.21 | |
| | | | | 0.1 | 313.53 | |

| Legend | |
|---|---|
| 🟨 | Given Refenece Parameter |
| 🟩 | Our Research Derived equations and Calculation |
| 🟦 | Establishing Cataloug data |
| ▨ | Experimental Reference |
| ░ | Converted to Reservoir Condition Field application |

## The Dual-End Point P-Function $P_c$ Curve Construction

After Eq. 8-10 validation, we build another $P_c$ point to produce a power function that represents the $P_c$ curve. That point would be another boundary point. We used the point at the ultimate low water saturation ($S_w$) ~zero but cannot be zero because it creates singularity for a power function with a limit of zero on the x - Axis. The power function has another boundary in the y-axis, which is ∞. We call this point the Ultimate Capillary Pressure ($P_{cu}$). Therefore, the points of selection can be in the range of $0.01 < S_w < 0.03$ in the x-Axis, while the pressure range is $7000 < P_c < 10000$Psi. The high-pressure value of 10000psi is just above the fracturing pressure of this research rock, which ensures that we are covering the whole possible ranges of pressure values for that rock. Therefore, a higher-pressure value can be used as needed for other formations with higher fracturing pressure. We used these two points, $P_{cd}$ and $P_{cu}$, to find the capillary pressure power function as a function of water saturation.

**The Dual-End Point P-Function Validation with C-Function**

We did additional validation for P-Function in using the two endpoints. We compare P-Function results with the conventional wisdom, the C-Function of the primary-drainage capillary curve, with evident matching as shown in Figure 1. We display the results of comparing C-Function and P-Function for two rock types. A high permeability rock type (blue) and a lower permeability rock type (orange). At the same time, the results of the P-Function are in grey for the high permeability rock and in light orange color for the lower permeability rock.

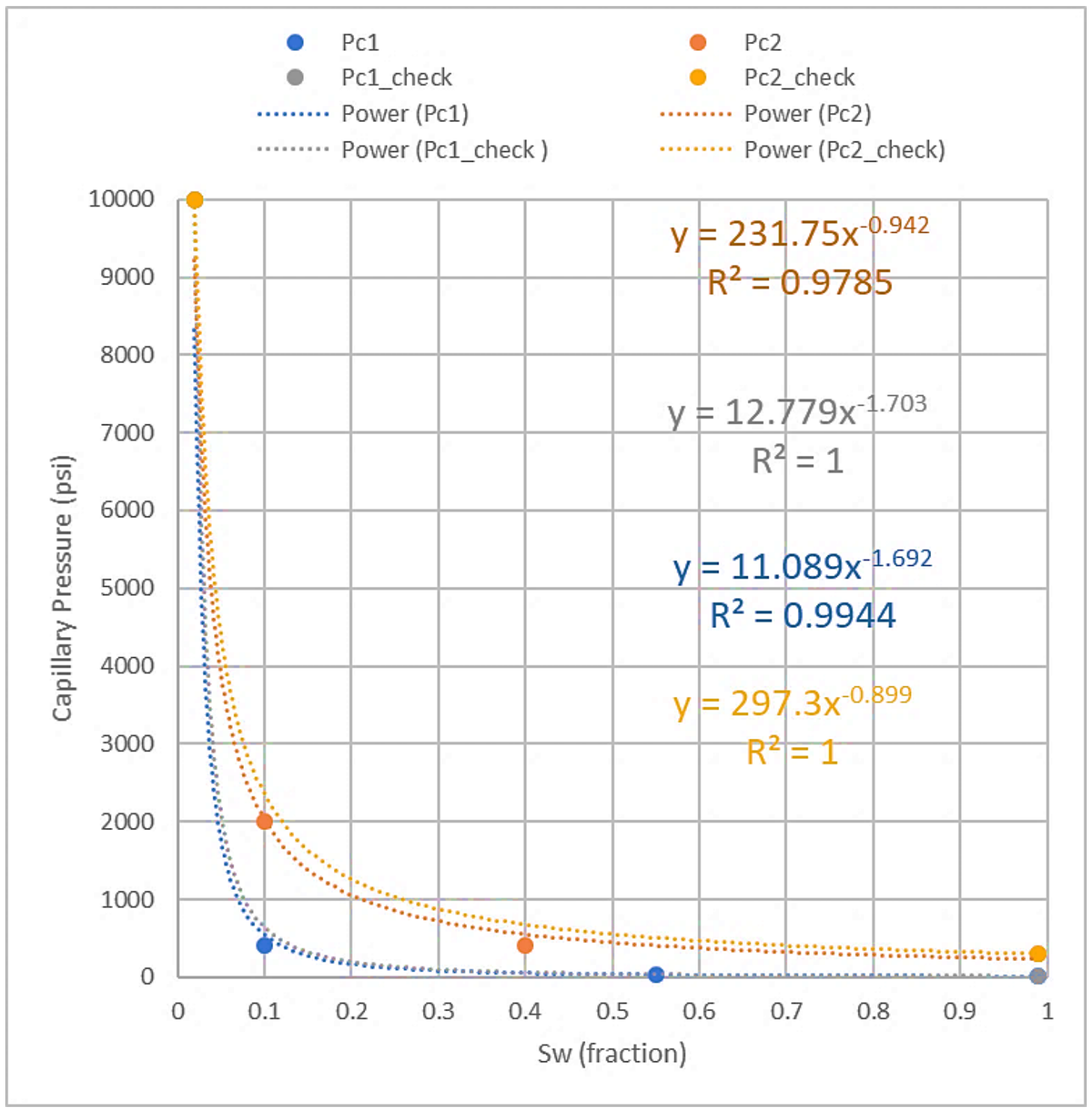

Figure 1. Validation of our novel P-Function (grey and light orange) with the conventional wisdom, the C-Function (blue and orange), respectively.

## Conclusions

We conclude that the two physical properties of porous media, capillary pressure, and permeability, have an intertwined behavior, with an inverse relationship. Both properties are also related to the pore throat network through the P-Function. Our research shows that if we know any of the four physical properties, grain size, pore throat size, capillary pressure, and permeability, all the others can be determined.

## Affiliation

Omar Alfarisi (ADNOC Offshore), Djamel Ouzzane (ADNOC), Mohamed Sassi (Khalifa University), and TieJun Zhang (Khalifa University).

## Acknowledgment

The authors thank the support and encouragement received from ADNOC, ADNOC Offshore, and Khalifa University of Science and technology. We express our appreciation to Mr. Yasser Al-Mazrouei, Mr. Ahmed Al-Suwaidi, Mr. Ahmed Al-Hendi, Mr. Ahmed Al-Riyami, Mr. Andreas Scheed, Mr. Hamdan Al-Hammadi, Mr. Khalil Ibrahim, Mr. Mohamed Abdelsalam, Dr. Ashraf Al-Khatib, Prof. Isam Janajreh, Dr. Aikifa Raza, Dr. Hongxia Li, and Mr. Hongtao Zhang.

## References

1. Y. Zhang *et al.*, Magnetic-actuated "capillary container" for versatile three-dimensional fluid interface manipulation. *Science Advances* **7**, eabi7498 (2021).
2. N. A. Dudukovic *et al.*, Cellular fluidics. *Nature* **595**, 58-65 (2021).
3. E. M. Landis, Capillary pressure and capillary permeability. *Physiological Reviews* **14**, 404-481 (1934).
4. E. C. Childs, N. Collis-George, The permeability of porous materials. *Proceedings of the Royal Society of London. Series A. Mathematical and Physical Sciences* **201**, 392-405 (1950).
5. W. Hassen *et al.*, Control of Magnetohydrodynamic Mixed Convection and Entropy Generation in a Porous Cavity by Using Double Rotating Cylinders and Curved Partition. *ACS Omega*, (2021).
6. H. B. Park, J. Kamcev, L. M. Robeson, M. Elimelech, B. D. Freeman, Maximizing the right stuff: The trade-off between membrane permeability and selectivity. *Science* **356**, (2017).
7. S. G. Ghedan, T. Weldu, O. Al-Farisi, in *Abu Dhabi International Petroleum Exhibition and Conference*. (Society of Petroleum Engineers, 2010).
8. T. W. Teklu, S. G. Ghedan, O. Al Farisi, in *SPE Production and Operations Conference and Exhibition*. (Society of Petroleum Engineers, 2010).
9. O. Al-Farisi *et al.*, in *Abu Dhabi International Conference and Exhibition*. (Society of Petroleum Engineers, 2004).
10. O. Alfarisi, Z. Aung, M. Sassi, Deducing of Optimal Machine Learning Algorithms for Heterogeneity. *arXiv preprint arXiv:2111.05558*, (2021).
11. M. Leverett, Capillary behavior in porous solids. *Transactions of the AIME* **142**, 152-169 (1941).
12. O. Al-Farisi, H. Belhaj, F. Yammahi, A. Al-Shemsi, H. Khemissa, in *International Conference on Offshore Mechanics and Arctic Engineering*. (American Society of Mechanical Engineers, 2013), vol. 55409, pp. V006T011A007.

13. O. Al-Farisi, M. Elhami, A. Al-Felasi, F. Yammahi, S. Ghedan, in *SPE/EAGE Reservoir Characterization & Simulation Conference*. (European Association of Geoscientists & Engineers, 2009), pp. cp-170-00073.
14. E. BinAbadat *et al.*, in *SPE Reservoir Characterisation and Simulation Conference and Exhibition*. (OnePetro, 2019).
15. D. Tiab, E. C. Donaldson, *Petrophysics: theory and practice of measuring reservoir rock and fluid transport properties*. (Gulf professional publishing, 2015).
16. O. Alfarisi *et al.*, Morphology Decoder: A Machine Learning Guided 3D Vision Quantifying Heterogenous Rock Permeability for Planetary Surveillance and Robotic Functions. *arXiv preprint arXiv:2111.13460*, (2021).
17. D. Beard, P. Weyl, Influence of texture on porosity and permeability of unconsolidated sand. *AAPG bulletin* **57**, 349-369 (1973).
18. B. Harrison, X. Jing, in *SPE Annual Technical Conference and Exhibition*. (OnePetro, 2001).